\documentclass[12pt]{article}
\pdfoutput=1
\usepackage[english]{babel}
\usepackage[utf8]{inputenc}
\usepackage[T1]{fontenc}
\usepackage{amsmath}
\usepackage{amsthm}
\usepackage{url}
\usepackage{authblk}
\usepackage{tensor}
\usepackage[bookmarks,colorlinks=false]{hyperref}
\usepackage{mathrsfs}

\newcounter{mnotecount}[section]

\newcommand{\beq}{\begin{eqnarray}}
\newcommand{\eeq}{\end{eqnarray}}
\newcommand{\ben}{\begin{eqnarray*}}
\newcommand{\een}{\end{eqnarray*}}
\newcommand{\Al}{A_\lambda}
\newcommand{\Bl}{B_\lambda}
\newcommand{\Fl}{F_\lambda}
\newcommand{\av}{\mathscr{A}}
\newcommand{\tv}{\mathscr{T}}

\newcommand{\tz}{\tilde{z}}

\newtheorem*{theorem*}{Theorem}

\DeclareMathOperator*{\wlim}{w-lim}

\begin{document}
\title{Backreaction for Einstein-Rosen waves coupled to a massless scalar field}
\author[1,3]{Sebastian J. Szybka}
\author[2,3]{Michał J. Wyrębowski}
\affil[1]{Astronomical Observatory, Jagiellonian University}
\affil[2]{Institute of Physics, Jagiellonian University}
\affil[3]{Copernicus Center for Interdisciplinary Studies}
\date{}
\maketitle{}
\begin{abstract}
We present a one-parameter family of exact solutions to Einstein's equations that may be used to study the nature of the Green-Wald backreaction framework. Our explicit example is a family of Einstein-Rosen waves coupled to a massless scalar field. This solution may be reinterpreted as a generalized three-torus polarized Gowdy cosmology with scalar and gravitational waves. We use it to illustrate essential properties of the Green-Wald approach. Among other things we show that within our model the Green-Wald framework uniquely determines backreaction for finite-size inhomogeneities on a predefined background. The results agree with those calculated in the Charach-Malin approach. In the vacuum limit, the Green-Wald, the Charach-Malin and the Isaacson methods imply identical backreaction, as expected.
\end{abstract}

\section{Introduction}

In the Green-Wald framework \cite{greenwald} backreaction is represented by the effective energy-momentum tensor. It describes the leading-order effects of small-scale inhomogeneities on the global structure of spacetime. The main result of \cite{greenwald} states that this effective energy-momentum tensor satisfies the weak energy condition and is traceless. Thus, the significant backreaction effect of small-scale inhomogeneities can be produced only by high-frequency gravitational waves and not by matter inhomogeneities. This result holds for all spacetimes satisfying the mathematical assumptions of Green and Wald \cite{greenwald} and, if applied to cosmology, implies that backreaction cannot mimic the cosmological constant. 
The fundamental question is whether the Green-Wald framework captures the effect of inhomogeneities properly and whether it may be applied to our Universe.
Recently, this problem was the subject of a discussion \cite{antiWald,angryWald,OstrowskiRoukemaGW,nolambdaGW}. 

The most straightforward approach to clarify the issue described above is to search for exact spacetimes to which the Green-Wald framework may be applied and for which it leads to nontrivial backreaction effects. The study of such an example is the main aim of our paper.

The theorems of Green and Wald \cite{greenwald} are based on mathematical assumptions that are not intuitive. The main assumption is the existence of a one-parameter family of solutions to Einstein's equations, $g_{ab}(\lambda)$, $\lambda>0$, that approaches the effective (background) spacetime in \mbox{a special} way as $\lambda\rightarrow 0$. It is not necessary to know this family in \mbox{an exact} form to predict the properties of the backreaction effect. However, one should provide examples of such families to explore the class of solutions to which the Green-Wald framework applies. Moreover, since there exist other similar frameworks, one may use these examples of one-parameter families of exact solutions to check the consistency of different methods. In this paper, we compare the Green-Wald framework to the Charach-Malin approach \cite{charachmalin} and, in the vacuum limit, to the Isaacson method \cite{isaacson1,isaacson2}. Although these methods do not differ much when it comes to practical calculations, they are based on different mathematical assumptions. The Charach-Malin and Isaacson methods do not involve `taking limits', and thus there are no doubts that they describe backreaction for finite-size inhomogeneities.\footnote{This aspect of the Green-Wald method has been recently analyzed in \cite{nolambdaGW}.} The essential advantage of the Green-Wald formalism over other similar approaches is the fact that it allows one to prove general theorems about the properties of backreaction. These theorems do not have their counterparts in other approaches. Therefore, it is interesting to study relations between the different frameworks. The exact solution presented in this paper provides such an opportunity.
 
Some time ago we\footnote{In collaboration with Krzysztof Głód and Alicja Konieczny.} presented a one-parameter family of exact solutions to Einstein's equations that was used to test the Green-Wald backreaction framework \cite{ourwm}. This was the first nonvacuum example of a family of exact solutions that satisfies the Green-Wald assumptions. The simplicity of this family makes it an interesting testbed for this formalism. On the other hand, the energy-momentum tensor in \cite{ourwm} is uniformly convergent as $\lambda\rightarrow 0$, so its weak limit\footnote{For a definition of {\it a weak limit} see \cite{greenwald}.} coincides with the ordinary limit---such matter inhomogeneities do not model the inhomogeneities in our Universe sufficiently well.\footnote{This was pointed out by Stephen Green and Robert Wald and suggested in the summary of the paper coauthored by us \cite{ourwm}. Recently similar remarks were presented in \cite{antiWald}.} Therefore, it would be instructive to find a nonvacuum example such that the weak limit of the energy-momentum tensor is nontrivial and the density contrast of the matter is high. We provide such an example in this paper. In our example, the energy-momentum tensor and the Ricci scalar are not uniformly convergent as $\lambda\rightarrow 0$. The density contrast of matter does not vanish and remains bounded, but the frequency of the density fluctuations blows up as the background spacetime is approached. We show that in our model this kind of inhomogeneities in the distribution of a massless scalar field do not contribute in the leading order to the backreaction effect in accord with what was proved in a general setting in \cite{greenwald}. {\it The bounded nonvanishing amplitude, high-frequency} matter inhomogeneities studied in our article provide an example of an intermediate behavior between  {\it decreasing amplitude, high-frequency} matter inhomogeneities studied in \cite{ourwm} and {\it unbounded amplitude, high-frequency} matter inhomogeneities\footnote{This classification of inhomogeneities is based on the behavior of the amplitude and frequency of fluctuations of matter density as the background spacetime is approached along the one-parameter family of solutions.}, an example of which has yet to be found.

There already exist four articles in which Green and Wald presented their method \cite{greenwald,nolambdaGW,greenwald2,greenwald3} and we refer the reader to those papers for mathematical details. However, 
we think that it could be instructive to review their framework from a different perspective.

\section{Averaging and the Green-Wald framework}

In this section we briefly review the Green-Wald framework. We restrict ourselves to the `leading-order analysis.' We keep the mathematical material to a minimum, however, for the sake of clarity it seems necessary to introduce some definitions. For brevity, sometimes the surnames Green and Wald will be abbreviated as GW. 

Let $M$ be a Lorentzian manifold and $L(M)$ a set of (at least) $C^2$ Lorentzian metrics on $M$. Following Green and Wald, one may introduce\footnote{For example, as presented in the Introduction in \cite{nolambdaGW}.} the notion of metrics with small-scale inhomogeneities. We denote this subset of $L(M)$ by $L_{inh}(M)$. Hence $L_{eff}:=L(M) \backslash L_{inh}(M)$ corresponds to all metrics without small-scale inhomogeneities. Any element of this set, if not indicated otherwise, will be denoted with $g^{(0)}$ and called an effective metric. By $g_\lambda\subset L_{inh}(M)$ we define a one-parameter family of metrics, such that $g^{(0)}=\lim_{\lambda\rightarrow 0^+}g_\lambda\in L_{eff}(M)$ (uniform convergence). 

It is usually assumed that the backreaction problem will be solved if an appropriate averaging procedure is found. This procedure---let us denote it by $\av:L_{inh}(M)\rightarrow L_{eff}(M)$---if applied to an inhomogeneous metric $g$, should give an averaged effective metric $g^{(0)}$, and thus $\av(g)=g^{(0)}$. Then, it is natural to expect that a well-defined $\av$ will satisfy some basic properties: it should be covariant and unique. 

Unfortunately, it seems that this kind of approach is not the best way to proceed in most physically interesting situations. For example, finding a correct $\av$ is not sufficient to solve the cosmological backreaction problem: one still needs `the true metric of the Universe'\footnote{Or, maybe better, `the metric that encapsulates all essential properties of the true metric of the Universe.'}---let us denote this hypothetical metric as $g_U\in L_{inh}(M)$\footnote{We assume here that $M$ corresponds to a manifold which may be used to construct a model of the Universe.}---to average it to find the effective metric. It is usually expected that $\av(g_U)=g_{RW}$, where $g_{RW}\in L_{eff}(M)$ stands for the Robertson-Walker metric. However, if we had known $g_U$, then most important problems would be solved and one would need the effective metric $g_{RW}$ only to simplify calculations. On the other hand, although knowing $\av$ alone does not solve the problem, it could be helpful. One might  apply $\av$ to some toy models, study how small-scale inhomogeneities alter the averaged effective metric and try to generalize the results to a cosmological context. 

The Green-Wald framework that we are going to study in this paper was not devised to work in the way described above. This approach is a generalized perturbation theory and as in every perturbative approach, i.e.\ the Isaacson high-frequency limit \cite{isaacson1,isaacson2}, full covariance is lost. Since it could be not clear what {\it a perturbative approach} may mean in the context of backreaction, we clarify this issue below. 

First, as in every perturbative approach, the effective spacetime must be assumed {\it a priori}. This is not a disadvantage of the framework, because the theorems of Green and Wald \cite{greenwald} hold for any background spacetime and arbitrary `perturbations' satisfying their mathematical assumptions. Thus, the properties of backreaction in this framework are known in general and a particular form of the background metric was not needed in the proofs. The GW method is in some sense an inverse approach to the one presented at the beginning of this section. If $\av$ is an arbitrary averaging procedure such that $\av(g)=g^{(0)}$, then $\av^{-1}(g^{(0)})\subset L(M)$ corresponds to the whole class of possibly inhomogeneous metrics that may be effectively represented by $g^{(0)}$. Similarly, in the GW approach $\av_{GW}^{-1}(g^{(0)})$  
corresponds to a set of one-parameter solutions to Einstein's equations $g_\lambda$ such that $\lim_{\lambda\rightarrow 0^+}g_\lambda=g^{(0)}$ (uniform convergence) and that satisfy the remaining assumptions of Green and Wald [(iii) and (iv) in \cite{greenwald}].
[Hereafter, the Green-Wald assumptions (i)-(iv) will be termed for brevity {\it the GW assumptions}.] Thus, the GW assumptions provide a precise definition of $\av_{GW}^{-1}(g^{(0)})$ which corresponds to a `cloud' of possibly inhomogeneous metrics that are in {\it a weak}\footnote{The precise meaning of this word follows from the GW assumptions \cite{greenwald}.} sense not far from some effective background $g^{(0)}$. In the GW framework essential information about inhomogeneities is encoded in first and second derivatives of metrics in $\av_{GW}^{-1}(g^{(0)})$. This idea is best known from the Isaacson approach.\footnote{The Isaacson approach was originally applicable only to vacuum spacetimes. It was reformulated in a mathematically rigorous way by Burnett \cite{burnett}. The Green-Wald framework \cite{greenwald} generalizes the Burnett result to nonvacuum spacetimes. Nonvacuum direct generalizations of the Isaacson formalism are also possible \cite{PodolskySvitek}.} 

How is backreaction defined in this kind of perturbative frameworks? The metrics in $\av_{GW}^{-1}(g^{(0)})$ and the corresponding energy-momentum tensors satisfy Einstein's equations. We would like to represent these inhomogeneous metrics by an effective metric $g^{(0)}$ and a smoothed energy-momentum tensor $T^{(0)}$.\footnote{Green and Wald provided a procedure to determine $T^{(0)}$.} However, there is no reason why both quantities should satisfy Einstein equations. We still may pretend that they do if we add an additional term, so $G(g^{(0)})=8\pi(T^{(0)}+t^{(0)})$. This term $t^{(0)}$ is called the effective energy-momentum tensor. It represents the effect of inhomogeneities on the background spacetime. So, as it possibly happens in cosmology, if one would naively construct a cosmological model with $g^{(0)}$, then one would discover that the corresponding energy-momentum tensor contains a strange contribution $t^{(0)}$. This problem is known as a `fitting problem' \cite{fitting}. One assumes {\it a priori} that Einstein's equations hold for the effective metric $g^{(0)}$,\footnote{In cosmology, this choice is based on a best fit to observational data.} and then asks whether the corresponding energy-momentum tensor contains some unexpected corrections in the form of $t^{(0)}$. We point out that this kind of inverse approach is usually adopted in modern cosmology, where philosophical prejudices supported by astronomical data were used to construct the standard cosmological model. With the new observations the `best fit' model is being improved, but one roughly knows the effective cosmological spacetime. The question is whether the parameters that define it have been interpreted correctly. In particular, is the energy-matter content what it seems or is a part of it an artifact of averaging?

We have mentioned at the beginning of this section that one should not see $\av_{GW}$ as a straightforward averaging procedure. Indeed, if $\av_{GW}$ is applied in this way, it acts on one-parameter families of metrics $g_\lambda$ which satisfy the GW assumptions, but its action is trivial: $\av_{GW}=\lim_{\lambda\rightarrow 0^+}$. It seems to be more appropriate to define the GW framework as a map $\tv: g_\lambda\rightarrow t^{(0)}$, where $g_\lambda$ satisfies the GW assumptions, and thus already encodes information about the effective metric $g^{(0)}=\lim_{\lambda\rightarrow 0^+}g_\lambda$.
This allows one to prove some general theorems about relations between inhomogeneous metrics $\av_{GW}^{-1}(g^{(0)})$ and the energy content of the effective spacetime with the averaged metric $g^{(0)}$ which represents all these inhomogeneous metrics \cite{greenwald}. Since there was no need to assume a particular form of the effective metric $g^{(0)}$, Green and Wald's results tell us about properties of backreaction in general. Using notation introduced above the main result of Green and Wald \cite{greenwald} may be formulated as follows. 

\begin{theorem*}{(by Green and Wald \cite{greenwald}, 2011)}\\
Let $g_\lambda$ be a one-parameter family of solutions to Einstein equations satisfying the GW assumptions. The effective energy-momentum tensor \mbox{$t^{(0)}:=\tv( g_\lambda)$} satisfies {\it the weak energy condition} and is traceless.
\end{theorem*}

This theorem implies that only gravitational radiation may contribute in the leading order to the backreaction effect, and thus in cosmology small-scale inhomogeneities cannot mimic a cosmological constant. 
This theorem also shows the advantage of the Green-Wald approach over the Isaacson-like nonvacuum extension of the high-frequency limit \cite{charachmalin,noonan} and justifies the need to introduce one-parameter families of metrics $g_\lambda$. In the original Isaacson approach (vacuum spacetimes) positivity of the effective energy density was shown only under an additional ansatz \cite{isaacson2}. The reason for that is a limited control over the coordinate dependence. Thus, the GW approach is not only a mathematically rigorous extension of the Isaacson approach, but there is a qualitative difference: within the GW framework one may study the nature of backreaction itself (not only backreaction in particular models). On the other hand, this kind of approach also has some disadvantages. Imagine that we know $g_U$ and we expect that it may be effectively represented by a $g_{RW}$ metric. It would not be easy to estimate the size (not nature, but size) of the backreaction effect within the GW approach.\footnote{Since gravitational waves have been detected \cite{LIGO1}, it is of great importance to estimate their cosmological backreaction effect. Although we have good reasons to assume that it is negligible, one may still insist on calculating its precise magnitude and this aim is hard to achieve within the GW framework alone.} One would need to use observational data to determine the appropriate member of a $g_{RW}$ class and find a family of exact $g_\lambda$ metrics such that $g_1=g_{U}$, $\lim_{\lambda\rightarrow 0}g_\lambda=g_{RW} $.\footnote{The parameter $\lambda$ may always be rescaled to obtain $g_U$ for $\lambda=1$.} In addition to that, there may exist in principle many such families\footnote{This is called a `path dependence' in \cite{antiWald}.}
 $g_\lambda$ 
so in such a case, for a given $g_{RW}$ and $g_U$, the effective energy-momentum tensor $t^{(0)}$ would not be uniquely specified.\footnote{We study this issue for a particular one-parameter family $g_\lambda$ in Subsection \ref{sec:unique}.} Thus, the power of the GW approach comes from the rigorous identification of a `cloud' of inhomogeneous metrics that are in some sense not far from the effective metric. This makes the general studies of the nature of backreaction possible. The price for this identification has the form of the GW assumptions about the existence of appropriate one-parameter families which limits the ability of the framework to provide quantitative predictions. We are not aware of any other solution of this identification problem which would give easy and precise control over coordinate dependence.\footnote{In the GW approach any one-parameter family of diffeomorphisms $\Psi_\lambda$ is allowed as a coordinate transformation provided that $\Psi_0=id$ (or more generally: any $\Psi_0$ that modifies the limit $\lim_{\lambda\rightarrow 0}g_\lambda$ trivially is allowed).}

Finally, we return to the fundamental question of whether the Green-Wald framework applies to our Universe. It seems not to be easy to identify elements of $\av_{GW}^{-1}(g_{RW})$ and it is not obvious if the `true metric of the Universe' $g_U$ belongs to this class.\footnote{One may construct toy models that in some sense exhibit the backreaction effects, but cannot be studied directly within the GW framework \cite{korzynski,helsinki}.} The most straightforward approach that may clarify this issue is to look for any exact families of metrics which belong to $\av_{GW}^{-1}(g^{(0)})$ for some effective metric $g^{(0)}$, which is not necessarily of physical interest. {\it A priori} there is no reason for the GW assumptions to be very restrictive\footnote{What is really restrictive is the fact that we are looking for solutions that are {\it exact} and, in addition, have a concise mathematical form.}; however, there are several logical possibilities to explore.\footnote{Since $t^{(0)}$ is traceless one should check if any solutions with nontrivial matter inhomogeneities may belong to $\av_{GW}^{-1}(g^{(0)})$. Otherwise, $t^{(0)}$ being traceless would be an artifact of assumptions that are too restrictive.} Any of the conditions described below would invalidate the GW framework as a proper tool to describe the backreaction effect of matter inhomogeneities:
\begin{enumerate}
\item  $\av_{GW}^{-1}(g^{(0)})=\emptyset$ for any $g^{(0)}$. \\

\vspace{-0.3cm}

It follows from examples presented in \cite{ourwm,greenwald2,burnett} that this is not the case.

\item  $\av_{GW}^{-1}(g^{(0)})$ contains only vacuum metrics for any $g^{(0)}$.\\

\vspace{-0.3cm}

This possibility was excluded by the example presented in \cite{ourwm}.

\item $\av_{GW}^{-1}(g^{(0)})$ contains only metrics with trivial matter inhomogeneities and gravitational waves on the $g^{(0)}$ background.

Such a possibility would naturally explain why the effective energy-momentum tensor is traceless. The example presented in this paper excludes this scenario.

\item $g_U\in\av_{GW}^{-1}(g_{RW})$ and $\exists g_\lambda$ satisfying the GW assumptions such that $g_1=g_U$, $\lim_{\lambda\rightarrow 0}g_\lambda=g_{RW}$  for the observational best-fit $g_{RW}$, but $\tv(g_\lambda)$ does not give the correct $t^{(0)}$.

\end{enumerate}

The possibilities $1$ and $2$ seem unnatural and they have already been excluded. The aim of this paper is to exclude the possibility $3$. In this article we present a one-parameter family of solutions to Einstein's equations $g_\lambda$ which satisfies the GW assumptions and which has nontrivial matter inhomogeneities. By the `nontrivial matter inhomogeneities' we understand inhomogeneities which are represented by the energy-momentum tensor which is not uniformly convergent in the limit $\lambda\rightarrow 0$.

\section{Einstein-Rosen waves and a massless\\ minimally coupled scalar field}\label{sec:2}

The cylindrically symmetric metric with two hypersurface orthogonal Killing fields $\partial_{\tz}$, $\partial_\varphi$ may be written in the form
\beq\label{metric}
g=e^{2(\gamma-\psi)}\left(-dt^2+d\rho^2\right)+\rho^2e^{-2\psi}d\varphi^2+e^{2\psi}d\tz^2\,,\label{met}
\eeq
where $\rho>0$, $-\infty<t,\tz<\infty$, $0\leq\varphi<2\pi$ and the metric functions $\psi$ and $\gamma$ depend on $t$ and $\rho$ only. Some nontrivial vacuum spacetimes with the metric \eqref{met} were discovered by Beck \cite{Beck25}, but they are better known as {\it Einstein-Rosen waves} \cite{EinsteinRosen37,Rosen54}. In this paper, we will investigate nonvacuum generalization of these solutions---a massless minimally coupled scalar field $\tilde\phi(t,\rho)$ will be added. 

There is an almost direct correspondence between generalized {\it Einstein-Rosen waves} and generalized Gowdy cosmologies. A subset of solutions studied in this paper may be reinterpreted as polarized three-torus Gowdy models. Thus, the example that we provide may be seen as a nonvacuum generalization of the Gowdy solution presented by Green and Wald in \cite{greenwald2} and a special case of the polarized Gowdy cosmologies investigated by Charach and Malin \cite{charachmalin}. Since Charach and Malin proposed their own approach to study the high-frequency limit of their solutions \cite{charachmalin}, we find it instructive to compare our results. This is done in Section \ref{sec:comparison}.

The energy-momentum tensor of a massless scalar field has the form
\ben
T_{ab}=\partial_a\tilde\phi\partial_b\tilde\phi-\frac{1}{2}g_{ab}\partial_c\tilde\phi\partial^c\tilde\phi\;.
\een
It is convenient to rescale the scalar field $\tilde\phi$, so from now on we will use $\phi$, where $\tilde\phi=\frac{1}{2\sqrt\pi}\phi$.
Einstein's equations
reduce to\footnote{In the following, dots and primes denote derivatives with respect to $t$ and $\rho$, respectively.}
\begin{align}
&\psi''+\frac{1}{\rho}\psi'-\ddot{\psi}=0\,,\label{psieq}\\
&\gamma'=\rho\left(\dot{\phi}^2+\phi'^2+\dot{\psi}^2+\psi'^2\right),\label{gampr}\\
&\dot{\gamma}=2\rho\left(\dot{\phi}\phi'+\dot{\psi}\psi'\right)\,.\label{gamdot}
\end{align}
The scalar field $\phi$ satisfies
\beq
&\phi''+\dfrac{1}{\rho}\phi'-\ddot{\phi}=0\,\label{fieq}
\eeq
which corresponds to $\nabla^a\nabla_a\phi=0$ and follows from the contracted Bianchi identities. The equations \eqref{psieq} and \eqref{fieq} are linear and identical to the equations for a polarized cylindrical wave propagating with the speed of light in Euclidean space. These equations are decoupled and may be solved separately. Once solutions to the equations \eqref{psieq} and \eqref{fieq} are known, the remaining metric function $\gamma$ may be found via the equations \eqref{gampr} and \eqref{gamdot} using quadratures. Moreover, the form of equations \eqref{gampr} and \eqref{gamdot} implies that every smooth pair of functions $\psi$, $\phi$ give rise to a smooth metric \eqref{met}. The Minkowski spacetime corresponds to $\psi$ and $\phi$ being constant. Decoupling of the dynamical equations for $\psi$ and $\phi$ implies that every solution of the vacuum equations ($\phi=0$) may be trivially extended to contain a nontrivial scalar field. [For a given $\psi$, it is sufficient to take any $\phi$ satisfying \eqref{fieq} and calculate the new function $\gamma$ via quadratures.]

In the succeeding section we will need the energy density of the scalar field $\epsilon$ as measured by the observers comoving with the coordinate system (with the four-velocity $u=e^{\psi-\gamma}\partial_t$)
\beq
\epsilon=T_{ab}u^au^b=\frac{1}{8\pi}e^{2(\psi-\gamma)}\left(\dot\phi^2+\phi'^2\right)\;,\label{emt}
\eeq
and the Ricci scalar $R$, which, with the help of the equations \eqref{psieq} and \eqref{fieq}, may be written in the form
\beq\label{RicciScalar1}
R=2e^{2(\psi-\gamma)}\left(\phi'^2-\dot\phi^2\right)\;.
\eeq

In general, spacetimes considered in this section may contain gravitational radiation. The analysis of backreaction of high-frequency gravitational waves in these spacetimes was presented by Podolský and Svítek \cite{PodolskySvitek}. 

\section{One-parameter family of solutions}

Our construction of a one-parameter family of metrics satisfying the assumptions of the Green-Wald framework is based on the solution described in the previous section. Namely, we choose the following particular solutions of the equations (\ref{fieq}) and (\ref{psieq}), respectively\footnote{One may consider a more general class of solutions, but we would like to make our example as simple as possible.}:
\beq\label{fpil}
\phi_{\lambda}(t,\rho)=\alpha\sqrt{\lambda}\; F_\lambda(t,\rho)\;,\quad\psi_{\lambda}(t,\rho)=\beta\sqrt{\lambda}\; F_\lambda(t,\rho)\;,\quad\lambda>0\;,
\eeq
where $\Fl(t,\rho)=J_0\left(\frac{\rho}{\lambda}\right)\sin\left(\frac{t}{\lambda}\right)$ and $\lambda$ is a parameter.
$J_0$ is the Bessel function of the first kind and zero order. The constants $\alpha$, $\beta$ are real and independent of $\lambda$. Integrating the equations \eqref{gampr} and \eqref{gamdot} and setting the additive integration constant to zero we obtain
\beq\label{gaml}
\gamma_{\lambda}(t,\rho)=
\;\frac{(\alpha^2+\beta^2)}{2\lambda}\rho^2\left[J_0^2(\frac{\rho}{\lambda})+J_1^2(\frac{\rho}{\lambda})
-2\frac{\lambda}{\rho} J_0(\frac{\rho}{\lambda})J_1(\frac{\rho}{\lambda})\sin^2(\frac{t}{\lambda})\right].
\eeq
Our one-parameter family of solutions to Einstein's equations, denoted $g_{ab}(\lambda)$ (where $\lambda>0$ is the parameter), has the form (\ref{met}) with the metric functions $\psi$, $\gamma$ and the scalar field given by \eqref{fpil} and \eqref{gaml}.

For $\rho/\lambda\gg 1$ the asymptotic behavior of the Bessel functions is given by
\beq\label{abes}
J_n(\frac\rho\lambda)=\sqrt{\frac{2}{\pi}\frac{\lambda}{\rho}}\left[\cos(\rho/\lambda-\frac{\pi}{2}n-\frac{\pi}{4})+O(\frac{\lambda}{\rho})\right]\;.
\eeq
Therefore, in the limit $\lambda\rightarrow 0$, we have 
\beq\label{limf}
\psi_{\lambda}\rightarrow 0\;,\quad \gamma_{\lambda}\rightarrow (\alpha^2+\beta^2)\rho/\pi\;,\quad \phi_{\lambda}\rightarrow 0\;.
\eeq
The background metric $g^{(0)}_{ab}:=\lim_{\lambda\rightarrow 0}g_{ab}(\lambda)$ is curved and has the form
\begin{equation}\label{metric0}
g^{(0)}=e^{2(\alpha^2+\beta^2)\rho/\pi}\left(-dt^2+d\rho^2\right)+\rho^2d\varphi^2+d\tz^2\,.
\end{equation}
The functions \eqref{limf} do not satisfy the equation \eqref{gampr}, and hence the metric $g^{(0)}$ does not belong to the class of solutions described in Section \ref{sec:2}.

Let $\Al(t,\rho)=J_0\left(\frac{\rho}{\lambda}\right)\cos\left(\frac{t}{\lambda}\right)$ and $\Bl(t,\rho)=J_1\left(\frac{\rho}{\lambda}\right)\sin\left(\frac{t}{\lambda}\right)$. Then for $\lambda>0$ the nonzero components of the energy-momentum tensor of the scalar field $\phi_\lambda$ take the form
\beq\label{Tsub}
T_{tt}(\lambda)&=&T_{\rho\rho}(\lambda)=\frac{\alpha^2}{8\pi\lambda}(\Al^2+\Bl^2)\;,\\\nonumber
T_{t\rho}(\lambda)&=&T_{\rho t}(\lambda)=-\frac{\alpha^2}{4\pi\lambda}\Al \Bl\;,\\\nonumber
T_{\varphi\varphi}(\lambda)&=&\frac{\alpha^2}{8\pi\lambda}e^{-2\gamma_{\lambda}}\rho^2(\Al^2-\Bl^2)\;,\\\nonumber
T_{\tz\tz}(\lambda)&=&T_{\varphi\varphi}(\lambda)\,\rho^{-2}e^{4\beta\sqrt{\lambda}\Fl}\;.
\eeq
The energy density of the scalar field measured by a comoving observer \eqref{emt} is
\beq\label{eden}
\epsilon(\lambda)=\frac{1}{8\pi}\frac{\alpha^2}{\lambda} e^{2(\beta\sqrt{\lambda}\Fl-\gamma_\lambda)}(\Al^2+\Bl^2)\;.
\eeq
The Ricci scalar \eqref{RicciScalar1} equals
\beq\label{RicciScalar2}
R(\lambda)=2\frac{\alpha^2}{\lambda} e^{2(\beta\sqrt{\lambda}\Fl-\gamma_\lambda)}(\Bl^2-\Al^2)\;.
\eeq 

Since the Bessel functions are regular at zero, it follows that the energy density and the Ricci scalar remain bounded as $\rho\rightarrow 0$ for any solution with nonzero $\lambda$.

In the formulas \eqref{Tsub}, \eqref{eden}, and \eqref{RicciScalar2}  the auxiliary functions $A_\lambda$, $B_\lambda$ may be approximated for a small value of $\lambda$ ($\rho/\lambda\gg 1$) with the help of \eqref{abes} by
\beq\label{Alim}
A_{\lambda}(t,\rho)&\approx&\sqrt{\frac{2}{\pi}\frac{\lambda}{\rho}}\cos\left(\frac{\rho}{\lambda}-\frac{\pi}{4}\right)\cos(\frac{t}{\lambda})\;,\\
\label{Blim}
B_{\lambda}(t,\rho)&\approx&\sqrt{\frac{2}{\pi}\frac{\lambda}{\rho}}\sin\left(\frac{\rho}{\lambda}-\frac{\pi}{4}\right)\sin(\frac{t}{\lambda})\;.
\eeq
In this approximation, the singular factor $1/\lambda$ in $T_{ab}$, $\epsilon$, $R$ cancels and these quantities exhibit rapid oscillatory behavior with a finite amplitude of oscillations\footnote{The frequency of oscillations blows up as $\lambda\rightarrow 0$, but the amplitude of oscillations is bounded and nonvanishing in this limit.}---the case which we refer to as {\it bounded nonvanishing amplitude, high-frequency} inhomogeneities.

\section{Inhomogeneity effect}

The equation satisfied by the background metric has the form \cite{greenwald}
\beq\label{eee}
G_{ab}(g^{(0)})=8\pi\,T_{ab}^{(0)}+8\pi\,t_{ab}^{(0)}\,,
\eeq
where $T_{ab}^{(0)}$ is so-called {\it the weak limit}\footnote{We use the symbol $\wlim$ to denote {\it the weak limit}.} of $T_{ab}(\lambda)$
$$T_{ab}^{(0)}=\wlim_{\lambda\rightarrow 0}T_{ab}(\lambda)\;.$$
The new term on the right-hand side, denoted as $t_{ab}^{(0)}$, is called the effective energy-momentum tensor and it encodes the backreaction effect of inhomogeneities. 
This tensor represents the additional terms that arise in the averaging process of Einstein's equations, so that these equations may still hold for $\lambda=0$ with the modified energy content.
A straightforward way to determine the effective energy-momentum tensor $t_{ab}^{(0)}$ is to calculate $t_{ab}^{(0)}=\frac{1}{8\pi}G_{ab}(g^{(0)})-T_{ab}^{(0)}$.

It follows from \eqref{Tsub}, \eqref{Alim}, and \eqref{Blim} that $T_{ab}(\lambda)$ is not uniformly convergent as $\lambda\rightarrow 0$. This is an essential novel property in comparison to the example published in \cite{ourwm}. In {\it the weak limit} the nonzero components of the scalar field energy-momentum tensor are
\ben
T_{tt}^{(0)}=T_{\rho\rho}^{(0)}=\frac{\alpha^2}{8\pi^2\rho}\;,
\een
so in the investigated model a fast-varying scalar field may be approximated by an anisotropic null fluid. Since the nonzero components of $G_{ab}(g^{(0)})$ are
\ben
G_{tt}(g^{(0)})=G_{\rho\rho}(g^{(0)})=\frac{\alpha^2+\beta^2}{\pi\rho}\;,
\een
then the nonzero components of the effective energy-momentum tensor are
\beq\label{eemt}
t_{tt}^{(0)}=t_{\rho\rho}^{(0)}=\frac{\beta^2}{8\pi^2\rho}\;.
\eeq
The effective energy-momentum tensor above is traceless and satisfies the weak energy condition, as predicted by theorems in \cite{greenwald}. It also has the form of an anisotropic null fluid\footnote{Similarly an energy-momentum tensor containing terms of the form of an anisotropic fluid appears in another approach to averaging \cite{coley_cosmo}.}
and depends on $\beta$, but does not depend on $\alpha$ (the parameter $\alpha$ controls the magnitude of the scalar field). Hence, inhomogeneities of the scalar field do not contribute in the leading order to the backreaction effect.

The Green-Wald framework requires the existence of a smooth tensor field\footnote{The derivative operator $\nabla_a$ is associated with the background metric $g_{ab}^{(0)}$.}
\ben
\mu_{abcdef}:=\wlim_{\lambda\rightarrow 0}\left[\nabla_ah_{cd}(\lambda)\nabla_bh_{ef}(\lambda)\right],
\een where $h_{ab}(\lambda):=g_{ab}(\lambda)-g_{ab}^{(0)}$. The effective energy-momentum tensor may be written in terms of $\mu_{abcdef}$
\begin{align}\nonumber
8\pi\,t_{ab}^{(0)}=\;\,&\frac{1}{8}\left(-\mu\indices{^c_c^{de}_{de}}-\mu\indices{^c_c^d_d^e_e}+2\,\mu\indices{^{cd}_c^e_{de}}\right)
g_{ab}^{(0)}+\frac{1}{2}\,\mu\indices{^{cd}_{acbd}}-\frac{1}{2}\,\mu\indices{^c_{ca}^d_{bd}}\\\label{eemtf}
&+\frac{1}{4}\,\mu\indices{_{ab}^{cd}_{cd}}-\frac{1}{2}\,\mu\indices{^c_{(ab)c}^d_d}+
\frac{3}{4}\,\mu\indices{^c_{cab}^d_d}-\frac{1}{2}\,\mu\indices{^{cd}_{abcd}}\;.
\end{align}
Although we have already calculated $t_{ab}^{(0)}$ from \eqref{eee}, it is instructive to determine $\mu_{abcdef}$ and check the consistency of the GW framework.
\mbox{A lengthy} calculation performed with the help of \textsc{Mathematica}\footnote{Some calculations were carried out with the {\sc xAct} package for {\sc Mathematica} \cite{xAct}.} yields
\begin{align*}
&\mu_{tttttt}=\mu_{tt\rho\rho\rho\rho}=-\mu_{tttt\rho\rho}\\
&=\mu_{\rho\rho tttt}=\mu_{\rho\rho\rho\rho\rho\rho}=-\mu_{\rho\rho tt\rho\rho}=\left[\frac{2}{\pi}\beta^2\rho^{-1}+\frac{1}{\pi^2}\left(\alpha^2+\beta^2\right)^2\right]e^{4(\alpha^2+\beta^2)\rho/\pi}\,,\\
&\mu_{tt\varphi\varphi\varphi\varphi}=\mu_{\rho\rho\varphi\varphi\varphi\varphi}=\frac{2}{\pi}\beta^2\rho^3\,,\\
&\mu_{tt\tz\tz\tz\tz}=\mu_{\rho\rho \tz\tz\tz\tz}=\frac{2}{\pi}\beta^2\rho^{-1}\,,\\
&\mu_{tt\rho\rho\varphi\varphi}=-\mu_{tttt\varphi\varphi}=\mu_{\rho\rho\rho\rho\varphi\varphi}=
-\mu_{\rho\rho tt\varphi\varphi}=\frac{2}{\pi}\beta^2\rho\, e^{2(\alpha^2+\beta^2)\rho/\pi}\,,\\
&\mu_{tt\rho\rho \tz\tz}=-\mu_{tttt\tz\tz}=\mu_{\rho\rho\rho\rho \tz\tz}=-\mu_{\rho\rho tt\tz\tz}=-\frac{2}{\pi}\beta^2\rho^{-1}e^{2(\alpha^2+\beta^2)\rho/\pi}\,,\\
&\mu_{tt\varphi\varphi \tz\tz}=\mu_{\rho\rho\varphi\varphi \tz\tz}=-\frac{2}{\pi}\beta^2\rho\,.
\end{align*}
All other components follow from symmetries \cite{greenwald}
\ben
\mu_{abcdef}=\mu_{(ab)(cd)(ef)}=\mu_{abefcd}
\een or are equal zero. We substitute this tensor into \eqref{eemtf} and calculate components of $t_{ab}^{(0)}$. They coincide with \eqref{eemt} in accord with the general calculations in \cite{greenwald}.

\section{Generalized three-torus Gowdy cosmologies}\label{sec:comparison}

The solutions studied in this paper may be reinterpreted as generalized Gowdy cosmologies. The main aim of this section is to compare our results to the high-frequency limit presented in \cite{charachmalin}.

Charach and Malin found a general solution which represents polarized Gowdy cosmologies with three-torus topology that are minimally coupled to a massless scalar field \cite{charachmalin}. Using their notation the metric may be put into the following form:
\begin{equation}\label{chmametric}
\hat g=L^2\left[e^f(-d\xi^2+dz^2)+\xi e^pdx^2+\xi e^{-p} dy^2\right]\;,
\end{equation}
where $L$ is a constant (for simplicity we assume from now on that $L=1$), $0\leq z<2\pi$, $\xi>0$, $0\leq x,y<2\pi$. The metric functions $f$, $p$ and the scalar field\footnote{The letter $\varphi$ denotes the angular variable in our original metric \eqref{metric} (not a scalar field), but we prefer to keep the original notation as in \cite{charachmalin}. Hopefully, this will not lead to confusion.} $\varphi$ depend only on $z$, $\xi$; periodicity in $z$ is assumed. The following formal complex substitution of variables and redefinition of metric functions brings the metric \eqref{metric} into the form: \eqref{chmametric}
\begin{eqnarray}\label{trans}
(t,\rho,\varphi,\tz)&\rightarrow&(i z,i \xi,i y,x)\;,\\\nonumber
\psi(t,\rho)&\rightarrow&\frac{1}{2}(p(z,\xi)+\ln\xi)\;,\\\nonumber
\gamma(t,\rho)&\rightarrow&\frac{1}{2}(f(z,\xi)+p(z,\xi)+\ln\xi)\;,\\\nonumber
\phi(t,\rho)&\rightarrow&2\sqrt{\pi}\varphi(z,\xi)\;.\nonumber
\end{eqnarray}
The variables $z$ and $\xi$ are assumed to be real, and thus one generates a new solution to Einstein's equations from the old one by a complex substitution. We note that the original metric \eqref{metric} was not periodic in $t$, but the metric \eqref{chmametric} is assumed to be periodic in $z$ with the period $2\pi$. In addition, it is assumed that $x$ is a periodic variable, but $\tz$ was not periodic for Einstein-Rosen waves. The original range of coordinates $t$, $\tz$ has changed. In three-torus Gowdy models they correspond to $z$ and $x$, respectively, and are periodic with the period $2\pi$. This implies that the cosmological solutions must be periodic in $z$, so the functions $\psi(t,\rho)$, $\gamma(t,\rho)$, $\phi(t,\rho)$ are substituted by periodic functions $p(z,\xi)=p(z+2\pi k_1,\xi)$, $f(z,\xi)=f(z+2\pi k_2,\xi)$, $\varphi(z,\xi)=\varphi(z+2\pi k_3,\xi)$, where $k_i$ are arbitrary integers. We also note that time and space coordinates are swapped: $\rho$ was a spatial coordinate, but $\xi$ is a time coordinate and $t$ was a time coordinate, but $z$ is a spatial coordinate.

The solutions \eqref{fpil} and \eqref{gaml} studied in this paper are related to a general solution discovered by Charach and Malin \cite{charachmalin}. Since Charach and Malin assumed periodicity in $z$ (three-torus cosmologies), and we did not assume a periodicity in $t$, this correspondence is not one to one. Let $j$ be a fixed large integer $j\gg 1$ and choose $\lambda=1/j$. (In some calculations $\lambda$ will be treated as a discrete parameter that approaches zero as $j\rightarrow+\infty$.) In order to see the correspondence, one should substitute (no summation over $j$ below) 
\begin{equation}\label{param}
\alpha_0=-1\;,\;\;A_n=\delta^j_{\;n}2\beta/\sqrt{j}\;,\;\; C_n=\frac{1}{2\sqrt{\pi}}\delta^j_{\;n}\alpha/\sqrt{j}\;,\;\;z_n=\frac{\pi}{2j}\delta^j_{\;n}\;
\end{equation} into solutions ($16$), ($17$), ($18$), and ($19$) in \cite{charachmalin}, set $j=1/\lambda$ and put all the remaining constants to zero. Finally, the inverse transformation to \eqref{trans} gives \eqref{fpil} and \eqref{gaml} from ($16$), ($17$), ($18$), and ($19$) in \cite{charachmalin}. [Note that the formula for the function $f_S$ was not described properly in \cite{charachmalin}. In order to obtain it from their equation ($19$), it is necessary to replace $\alpha_0$ by $\beta_0$, multiply all terms by a factor $16\pi$ and replace $A_n$ by $C_n$ and $B_n$ by $D_n$.]

Using the procedure above we find that our solution after the substitution \eqref{trans} corresponds to the metric \eqref{chmametric} with
\begin{eqnarray}\label{Gowdymetric}\nonumber
p&=&-\ln{\xi}+2\beta\sqrt{\lambda}J_0(\frac{\xi}{\lambda})\sin(\frac{z}{\lambda})\;,\quad
\varphi\:\,=\:\,\frac{1}{2\sqrt{\pi}}\alpha\sqrt{\lambda}J_0(\frac{\xi}{\lambda})\sin(\frac{z}{\lambda})\;,\\\nonumber
f&=&\frac{\alpha^2+\beta^2}{\lambda}\xi^2\left[J_0^2(\frac{\xi}{\lambda})+J_1^2(\frac{\xi}{\lambda})-2\frac{\lambda}{\xi}J_0(\frac{\xi}{\lambda})J_1(\frac{\xi}{\lambda})\sin^2(\frac{z}{\lambda})\right]\\
&&-2\beta\sqrt\lambda J_0(\frac\xi\lambda)\sin(\frac{z}{\lambda})\;,
\end{eqnarray}
where $\varphi$ denotes the scalar field. 

\subsection{The high-frequency limit}

It is instructive to calculate the high-frequency limit of our solution using the Charach-Malin procedure and compare it to the high-frequency limit in the Green-Wald approach. 

Using \eqref{abes} for $\xi/\lambda\gg 1$ we obtain to leading order in $\lambda/\xi$
\begin{eqnarray}\label{eq:hf1}
p&\approx&-\ln\xi +\frac{2\sqrt{2}}{\sqrt{\pi}}\frac{\beta\lambda}{\sqrt{\xi}}\cos\left(\frac{\xi}{\lambda}-\frac{\pi}{4}\right)\sin(\frac z \lambda)\;,\\\nonumber
\varphi&\approx&\frac{1}{\sqrt{2}\pi}\frac{\alpha\lambda}{\sqrt{\xi}}\cos\left(\frac{\xi}{\lambda}-\frac{\pi}{4}\right)\sin(\frac z \lambda)\;,\\\nonumber
f&\approx&\frac{2}{\pi}(\alpha^2+\beta^2)\xi\;,
\end{eqnarray}
in accord with the formulas ($34$), ($35$) and ($36$) in \cite{charachmalin} (with our choice of constants). Following Charach and Malin we decompose the metric \eqref{chmametric} into the `background' $\eta$ and `wave' part $h$ as $\hat g\approx\eta+h$, where
\begin{eqnarray}\label{eq:hf2}
\eta&=&e^{2(\alpha^2+\beta^2)\xi/\pi}(-d\xi^2+dz^2)+dx^2+\xi^2dy^2\;,\\\nonumber
h&=&\overline p(dx^2-\xi^2dy^2)\;,\\\nonumber
\overline p&=&p+\ln\xi\;.
\end{eqnarray}
We assumed that $\overline p$ is small. After appropriate redefinition of variables and functions Charach and Malin's background metric $\eta$ corresponds to the Green-Wald background metric $g^{(0)}$ given by \eqref{metric0}.

In order to show that the background geometry is created partly by the gravitational waves and partly by the scalar field Charach and Malin calculated the energy-momentum tensor of the background metric $\eta$ and decomposed it into the traceless part $T^{(1)}$ and nontraceless part $T^{(2)}$; hence \mbox{$\hat T=T^{(1)}+T^{(2)}=\frac{1}{8\pi}G(\eta)$}. For our solution we find
\begin{equation}
\hat T=T^{(1)}=\frac{\alpha^2+\beta^2}{8\pi^2\xi}(d\xi^2+dz^2)\;,\;\;\;T^{(2)}=0\;,
\end{equation}
where the nontraceless part $T^{(2)}$ vanishes because of our choice $\beta_0=0$ ($\beta_0$ is one of the constants in the full Charach-Malin solution). The traceless part $T^{(1)}$ is further decomposed\footnote{This decomposition is based on the interpretation of constants, e.g. for our solution $\alpha$ controls the amplitude of the scalar field $\varphi$.} into the gravitational waves part $T^{GW}$ and the scalar radiation part $T^{SW}$
\begin{equation*}
T^{(1)}=T^{GW}+T^{SW}\;,\;\;\;T^{GW}=\frac{\beta^2}{8\pi^2\xi}(d\xi^2+dz^2)\;,\;\;\;T^{SW}=\frac{\alpha^2}{8\pi^2\xi}(d\xi^2+dz^2)\;.
\end{equation*}
Let $k^\mu$ be a null vector defined by its covariant components as $$k=\frac{1}{\sqrt{\pi\xi}}(d\xi+dx)\;.$$ Then, $8\pi T^{GW}_{\mu\nu}=\beta^2k_\mu k_\nu$ and $8\pi T^{SW}_{\mu\nu}=\alpha^2k_\mu k_\nu$. Thus, Charach and Malin interpreted both of those terms in the traceless part of the energy-momentum tensor as a null fluid representation of collisionless flows of `gravitons' and scalar massless particles. This is one of the main results in their paper \cite{charachmalin}. To sum up, we started with an inhomogeneous cosmological model with small-scale matter and `gravitational field' inhomogeneities and showed how to represent this model by an effective spatially homogeneous (but anisotropic) spacetime filled with collisionless flows of massless scalar particles and `gravitons.' 

Up to this point calculations in the Charach-Malin framework were almost identical to those in the Green-Wald approach, so it is a good moment to summarize the differences. First, Charach and Malin split the metric into $\eta$ and $h$ using their intuition and the interpretation of the metric $\eta$. In fact, this is done via `inspection' of the line element. In the coordinate system that they have chosen, the form of the line element has a natural split into the `background' and the `wave' part. This gauge-dependent intuition is strongly supported by a nice physical interpretation of $\hat T=\frac{1}{8\pi}G(\eta)$; {\it a priori} there is no reason why $\hat T$ calculated in such a way should correspond to any physically reasonable form of the energy-momentum tensor. In the Green-Wald approach $g^{(0)}$ is calculated as the limit $\lambda\rightarrow 0$ of the one-parameter family of metrics $g(\lambda)$. Similarly to the Charach-Malin approach, we must define in advance our background metric. Both backgrounds coincide because the coefficients $A_n$ and $C_n$ that we have chosen approach zero in an appropriate way in the limit $\lambda\rightarrow 0$.  We are interested in the backreaction effect of finite-size inhomogeneities (for a finite $0<\lambda_0\ll 1$) and up to this point there would be nothing `ultra-local' in our calculations if we had conducted them in the Green-Wald framework. The result would not depend so far on the choice of a one-parameter family of metrics along which the limit $\lambda=0$ is achieved, provided that those families coincide for $\lambda\rightarrow 0$ and $\lambda=\lambda_0$.\footnote{Using terminology introduced in \cite{antiWald}, there is no path dependence so far. In Section \ref{sec:unique} we will show that the final results are also path independent.}

In their paper, Charach and Malin did not put any restrictions on the choice of the coefficients $A_n$, $B_n$, $C_n$, $D_n$. It is rather obvious that they must be in some sense small: the wave part $h$ must be a small correction to the background $\eta$ (e.g., $\overline p$ is explicitly assumed to be small). Otherwise their result and its interpretation is not valid. In the Green-Wald approach this assumption is incorporated into the framework. The introduction of $\lambda$-dependent family of metrics allows us to define the split of the metric into the background $g^{(0)}$ and the `perturbation' $h$ in a gauge-independent manner (provided that transformations of coordinates do not alter the limit $\lambda\rightarrow 0$). 

In addition to the reasoning outlined above, Charach and Malin suggested an alternative procedure to show the null fluid type high-frequency behavior of the scalar field source. We may average the energy-momentum tensor. Namely, we take the energy-momentum tensor \eqref{Tsub} and apply the transformation \eqref{trans}. Next, we use asymptotic formulas for Bessel functions for $\xi/\lambda\gg 1$. Averaging over the phase the nontrivial components of the energy-momentum tensor gives 
\begin{eqnarray}\label{avTi}\nonumber
&&<T_{\xi\xi}>=<T_{zz}>=\frac{\alpha^2}{8\pi^2\xi}(1+\frac{1}{8}\frac{\lambda^2}{\xi^2}+\dots)\;,\\
&&<T_{xx}>=\frac{1}{\xi^2} <T_{yy}>=\frac{\alpha^2}{64\pi^3}e^{-2(\alpha^2+\beta^2)\xi/\pi}\left(\frac{2\pi}{\xi}+\alpha^2+\beta^2\right)\frac{\lambda^2}{\xi^2}+\dots\hspace{1cm}\;.\nonumber
\end{eqnarray}
In the leading order 
we have
\begin{equation}\label{avT}
<T_{\xi\xi}>=<T_{zz}>=\frac{\alpha^2}{8\pi^2\xi}\;,\;\;<tr(T)>=<T^x_{\;x}>=<T^y_{\;y}>=0\;,
\end{equation}
with the remaining components equal to zero and in accord with Charach and Malin's asymptotic averages: the equations ($62$), ($63$), and ($64$) in \cite{charachmalin}. Of course, this is the averaging of components of a tensor, so it is not a covariant procedure. The result is invariant against a restricted class of coordinate transformation, but this class has not been specified in \cite{charachmalin}. In order to overcome this kind of difficulties, Green and Wald averaged the energy-momentum tensor differently, namely, they took {\it a weak limit} of $T$ (we denote it as $T^{(0)}=\wlim_{\lambda\rightarrow 0}T$). Thus, in their calculation one `goes with $\lambda$ to zero' along a decreasing sequence of small numbers $1/m$, where $m=j,j+1,\dots,+\infty$. Such a procedure is needed only to average the energy-momentum tensor $T$ in a more gauge controlled way than in the Charach-Malin approach which has been demonstrated above. In the case under investigation the Green-Wald weak limit coincides with \eqref{avT}, and hence in the leading order $<T>=T^{(0)}$.

Charach and Malin did not discuss backreaction in their model; however, it seems reasonable to follow the Green-Wald approach and define backreaction in the Charach-Malin approach by the effective energy-momentum tensor $\hat t=\hat T-<T>$. Thus, in the Green-Wald and Charach-Malin approaches the beckreaction effect for the particular choice of solution studied in this paper is identical. It is given by the effective energy-momentum tensor which is traceless and corresponds to gravitational radiation 
\begin{equation}\label{eq:tGowdy}
\hat t=\frac{\beta^2}{8\pi^2\xi}(d\xi^2+dz^2)=t^{(0)}\;,
\end{equation}
where $t^{(0)}$ denotes the effective energy-momentum tensor in the GW approach.

\subsection{Uniqueness of the effective energy-momentum \mbox{tensor}}\label{sec:unique}

An interesting question arises: had we chosen $A_j$, $C_j$ differently, would the GW framework lead to a different effective energy-momentum tensor \eqref{eq:tGowdy}? Let us assume for a moment that $g^{(0)}=\eta$ and $g(\lambda_0)=\hat g$ for some large natural number $j=1/\lambda_0$. Physical inhomogeneities are always of finite size, so it seems that both conditions should define backreaction in the model uniquely. However, there may exist many one-parameter families $g(\lambda)$ that satisfy these conditions. The Green-Wald framework provides a mapping $\tv: g_\lambda\rightarrow t^{(0)}$ as described in the Introduction, and thus backreaction for different one-parameter families may differ. In \cite{antiWald} this was called a `path dependence.' We show below that in the model under investigation the results are `path independent.'

Let us consider modified solutions that differ from \eqref{param} in the choice 
\begin{equation}\label{altsol}
A_n=\delta^j_{\;n}2\beta/\sqrt{j^{\kappa/\sigma}}\;,\;\;\; C_n=\frac{1}{2\sqrt{\pi}}\delta^j_{\;n}\alpha/\sqrt{j^{\iota/\sigma}}\;,\;\;\;j=\frac{1}{\lambda^\sigma}\;.
\end{equation} 
It follows from the definitions above that $A_j\sim \lambda^{\kappa/2}$ and $C_j\sim \lambda^{\iota/2}$. Of course, other choices are also possible, but we would like to have $g^{(0)}=\eta$, \mbox{$g(\lambda_0)=\hat g$} and since $j$ is assumed to be large, we are interested in the asymptotic behavior of $A_j$ and $C_j$; thus \eqref{altsol} seems to be a reasonable parametrization.\footnote{In fact, the analysis presented in this section covers a wide class of $\lambda$-dependent families of solutions $g(\lambda)$ for which the leading terms of some general functions $A_n(\lambda)$, $C_n(\lambda)$, $j(\lambda)$ for small $\lambda$ have the form \eqref{altsol}.} The solutions depend now on three more constants $\kappa$, $\iota$ and $\sigma$. The constant $\sigma$  must be positive because otherwise $\lambda\rightarrow 0$ would not correspond to the high-frequency limit. The leading term in the formula for $f$ has the form
\begin{equation}
f\approx\frac{2}{\pi}\left(\alpha^2\lambda^{\iota-\sigma}+\beta^2\lambda^{\kappa-\sigma}\right)\xi\;.
\end{equation} 
In order to recover the background $\eta$ in the limit $\lambda\rightarrow 0$ we must have $\iota-\sigma=\kappa-\sigma=0$; thus $\iota=\kappa=\sigma$ and there is only one independent parameter $\sigma$. However $\hat T$, $T^{(0)}=\wlim_{\lambda\rightarrow 0}T$ and the leading terms in \mbox{$<T>$} do not depend on $\sigma$.\footnote{Although $A_j$, $C_j$ and $j$ depend on $\sigma$ this dependency cancels in appropriate asymptotic expressions for $\hat T$, $<T>$, $T^{(0)}$ by the same mechanism as in the limit $\eta:=\lim_{\lambda\rightarrow 0} g_\lambda$, where the $\lambda$ approaches zero along a discrete sequence of values.} Thus, without loss of generality we may set $\sigma=1$ to recover solutions studied in this paper. To sum up, although there may exist many one-parameter families $g_\lambda$ with the property $g^{(0)}=\eta$, \mbox{$g(\lambda_0)=\hat g$} the Green-Wald framework in the model under investigation provides a unique procedure to calculate backreaction effects. The condition $\eta=g^{(0)}:=\lim_{\lambda\rightarrow 0}g_\lambda$ fixes uniquely the asymptotics in $\lambda$ of $g_\lambda$ which determines the effective energy-momentum tensor $t^{(0)}=\hat T-T^{(0)}$.\footnote{Since $\eta=g^{(0)}$, then $\hat T=\frac{1}{8\pi}G(g^{(0)})$.}
 
\subsection{More general solutions}

The crucial difference between the Charach-Malin and the Green-Wald approaches is illustrated by the following fact. In both approaches the effective energy-momentum tensor is traceless. In the Charach-Malin approach this is rather an `interesting result of calculations' that were conducted for a particular solution to Einstein's equations. In the Green-Wald approach this is a consequence of their theorems (as explained in the Introduction), and thus a fundamental property of all effective energy-momentum tensors that may be calculated within their framework.

In order to illustrate this property one may consider more general solutions than those studied in this paper, namely, we may take the Charach-Malin solution \cite{charachmalin} with the constant $\beta_0\neq 0$ and the remaining constants corresponding to those studied in this paper. This solution contains a spatially homogeneous component of the scalar field and it is similar in some sense to the Belinskii-Khalatnikov solution. The equation ($46$) in \cite{charachmalin} implies that for $\beta_0\neq 0$ the energy-momentum tensor $\hat T$ is not traceless $[tr(\hat T)=\beta_0^2\xi^{-2}e^{-f}]$ and one may try to use this fact to construct a counterexample to the Green and Wald theorems. The equation ($64$) in \cite{charachmalin} shows that for $\xi\gg 1$ the average energy-momentum tensor $<T>$ is traceless up to the order $\xi^{-1}$. This is consistent with the fact that the trace of the effective energy-momentum tensor $\hat t=\hat T-<T>$ should vanish in the highest order (terms $\xi^{-1}$) and remains in agreement with the Green and Wald theorems provided that we choose $\beta_0\sim \lambda$. If $\beta_0$ will not be proportional to $\lambda$, then the trace of $\hat T$ may contribute to the trace of the effective energy-momentum tensor $\hat t$. Thus, the Green and Wald theorems imply that there should appear a nonzero trace in the leading order of $<T>$ to cancel the trace of $\hat T$ and make $\hat t$ traceless. Our analogy between the Green-Wald and Charach-Malin approaches suggests that also $T^{(0)}$ cannot be traceless. Indeed, calculations show that in the leading order $tr(\hat T)=tr(<T>)=tr(T^{(0)})$ as predicted by Green and Wald and $tr(\hat t)=tr(t^{(0)})=0$ also for $\beta_0\neq 0$ in agreement with their theorems.

\subsection{Vacuum limit: The Isaacson approach}

The solution \eqref{Gowdymetric} corresponds for $\alpha=0$ to vacuum three-torus polarized Gowdy cosmologies.\footnote{The full metric $\hat g$, which is given by \eqref{chmametric}, is a vacuum solution, but the background metric $\eta$ is nonvacuum. Also $\eta+h$ does not satisfy the vacuum Einstein's equations exactly.} These cosmologies were already studied in a different coordinate system in the context of the GW framework in \cite{greenwald2}. The results presented there are consistent with ours, but the direct comparison is obscured by nontrivial coordinate transformation (the tensor $\mu_{abcdef}$ was not explicitly given in \cite{greenwald2}). The vacuum case $\alpha=0$ may be investigated within the Isaacson approach \cite{isaacson1,isaacson2}. 

We find it instructive to apply the Isaacson framework to the solutions studied in this paper. It follows from \eqref{eq:hf1} and \eqref{eq:hf2} that our metric \eqref{chmametric} may be written in the Isaacson form $\tilde g=\eta+\lambda \tilde h$, where $\tilde h=h/\lambda$. We have $\tilde h_{ab}=O(\lambda^0)$, $\nabla_{a} \tilde h_{bc}=O(\lambda^{-1})$ and $\nabla_{a}\nabla_{b} h_{cd}=O(\lambda^{-2})$, where $\nabla_a$ is a covariant derivative associated with $\eta$.
In the Isaacson approach the Ricci tensor is calculated in terms of $\tilde h$ and its derivatives. This expression is expanded in orders of $\lambda$ $$R_{ab}[\tilde g(\lambda)]=R_{ab}^{(0)}(\lambda)+\lambda R_{ab}^{(1)}(\lambda)+\lambda^2 R_{ab}^{(2)}(\lambda)+O(\lambda)\;.$$
We have $\lambda R_{ab}^{(1)}(\lambda)=O(\lambda^{-1})$ and the remaining two terms $R_{ab}^{(0)}(\lambda)$, $\lambda^2 R_{ab}^{(2)}(\lambda)$ are of the same order $O(\lambda^0)$. The vacuum Einstein's equations $R_{ab}[\tilde g(\lambda)]=0$ imply that $R_{ab}^{(1)}(\lambda)=0$ and $R_{ab}^{(0)}(\lambda)+\lambda^2 R_{ab}^{(2)}(\lambda)=0$; thus the average of the $R_{ab}^{(2)}$ may play the role of the effective energy-momentum tensor for the background metric [$R_{ab}^{(0)}(\lambda)=R_{ab}(\eta)$]. Therefore, following Isaacson \cite{isaacson2} and Brill and Hartle \cite{BH} (see also \cite{burnett}) we define $\tilde{t}_{ab}=-\frac{\lambda^2}{8\pi}<R_{ab}^{(2)}(\lambda)>$. In order to calculate it one may use equation ($2.8$) in \cite{isaacson1}, but it is more convenient for us to expand the final expression for $R_{ab}(\lambda)$ in powers of $\lambda$, and indentify $R_{ab}^{(0)}(\lambda)$ as $R_{ab}(\eta)$; the remaining terms in the order $O(\lambda^0)$ will give us $-\frac{1}{8\pi}\tilde t$. We find $\tilde t=\hat t=t^{(0)}$ in accord with \eqref{eq:tGowdy}. Therefore, all three frameworks (by Isaacson, by Charach and Malin, and by Green and Wald) predict the same backreaction in the vacuum limit. 

\section{Summary}

 The one-parameter family of exact nonvacuum solutions to Einstein's equations presented in this article satisfies all assumptions of the Green-Wald framework \cite{greenwald}. The other three examples of such families presented in the literature so far were restricted to vacuum \cite{greenwald2,burnett} or to stiff-fluid spacetimes \cite{ourwm}. In the cosmological context, the most interesting are nonvacuum families. The novel important property of the example presented in this article (in comparison to the nonvacuum example presented in \cite{ourwm}) is the nontrivial behavior of the energy-momentum tensor. The matter density exhibits {\it bounded nonvanishing amplitude, high-frequency} oscillations. Our calculations confirm the mathematical consistency of the Green-Wald approach. We showed that within the model studied it predicts the backreaction effects uniquely for a finite size of inhomogeneities. We used the family of exact solutions in question to compare the Green-Wald framework to the Charach-Malin approach, and in the vacuum limit to the Isaacson method. Although these methods use slightly different mathematical formalisms, all of them agree in their range of applications.

A task for the future is to find an example of a one-parameter family of solutions to Einstein's equations\footnote{Solutions that are preferably coupled to {\it a massive} field.} for which the amplitude of fluctuations of matter density becomes unbounded as the background spacetime is approached.

\vspace{0.5cm}

\noindent{\sc Acknowledgments}

\vspace{0.2cm}

We thank Alan Coley, Krzysztof G\l{\'o}d, and Szymon Sikora for comments and an anonymous referee for drawing our attention to the article by Charach and Malin \cite{charachmalin}. M. W. was partially supported by the Copernicus Center for Interdisciplinary Studies in Krak{\'o}w and the Foundation ORLEN -- DAR SERCA.

\bibliographystyle{unsrt}
\bibliography{report}

\end{document}